# Dynamical Ensembles in Nonequilibrium Statistical Mechanics

G. Gallavotti[1], E. G. D. Cohen[2]

[1]*Fisica, Università di Roma, "La Sapienza", 00185 Roma, Italia*

[2]*The Rockefeller University, New York, NY 10021, USA*

(October 30, 1994)

## Abstract

Ruelle's principle for turbulence leading to what is usually called the Sinai-Ruelle-Bowen distribution (SRB) is applied to the statistical mechanics of many particle systems in nonequilibrium stationary states. A specific prediction, obtained without the need to construct explicitly the SRB itself, is shown to be in agreement with a recent computer experiment on a strongly sheared fluid. This presents the first test of the principle on a many particle system far from equilibrium. A possible application to fluid mechanics is also discussed.
47.52, 05.45, 47.70, 05.70.L, 05.20, 03.20

Typeset using REVTEX

In his treatment of strange attractors to understand turbulence Ruelle introduced, as a principle (R): *the time averages of observables, on motions with initial data randomly sampled with the Liouville distribution $\bar\mu_0$, are described by a stationary probability distribution $\bar\mu$ obtained by attributing a suitable probability density to the surface elements of the unstable manifolds of the points in phase space*, [27]. In hyperbolic systems this leads, as a theorem, to identifying $\bar\mu$ with the SRB on the attractor, [27]. (R) was based on important previous results [29], [10], [26], [25], [17].

The unstable manifolds are quite difficult to obtain: it has been questioned whether (R) has any predictive value when it cannot be a priori proved. [1] It is implicit, we think, in Ruelle's ideas that the principle should be applied by assuming suitable properties making (R) valid and then retain the consequences concerning the large scale (in time and size) behavior. A similar situation arises in statistical mechanics where the equilibrium statistics is accepted on the basis of the ergodic principle (usually called hypothesis) and one derives the "heat theorem" (second law) and the thermodynamic properties. Analogous consequences, which would turn (R) into a tool for predictions, proved very hard to find in the case of turbulence. But recently [11] have shown that (R) does imply macroscopic consequences, e.g. the Einstein relation for the diffusion and conductivity coefficients in systems similar to ours in the linear regime where (R) can even be proven for a single particle system.

In this letter we present consequences of a different nature, that can be derived from (R) and subjected to experimental tests, particularly for *non small* forcing, *without* having to compute rather difficult dynamical quantities like Lyapunov exponents as in [11] and *without* the need to compute, even approximately, $\bar\mu$. [2]

Although our considerations appear to be quite general, the letter is written with the theoretical interpretation of a numerical experiment in mind: the entropy production fluctuations in a shearing many particle fluid in a nonequilibrium stationary state far from equilibrium, [15].

We first define the model of the shearing fluid we treat. The two-dimensional shearing fluid is in a non equilibrium stationary state, driven by an external shear rate $\gamma = \partial u_x/\partial y$ ("forcing") where **u** is the horizontal local velocity $\mathbf{u} = \mathbf{i}\gamma y$ (and **i** is the $x$ axis unit vector) and coupled to a thermostat. The equations of motion are the SLLOD equations [16], [14]:

$$\dot{\mathbf{q}}_j = \mathbf{p}_j/m + \mathbf{i}\gamma y_j; \qquad \dot{\mathbf{p}}_j = \mathbf{F}_j - \mathbf{i}\gamma p_{yj} - \alpha \mathbf{p}_j \qquad (1)$$

where $j = 1, ..., N$ labels the $N$ fluid particles with mass $m$, $\mathbf{p}_j/m$ is the peculiar velocity of particle $j$, i.e. its velocity with respect to the local fluid velocity $\mathbf{u}(\mathbf{q}_j) = \mathbf{i}\gamma y_j$; $\mathbf{F}_j$ is the interparticle force on particle $j$, due to a purely repulsive pair potential $\varphi(r)$ where $r$ is the interparticle distance (e.g. an inverse power potential). If $\mathbf{x} = (\mathbf{p}_1 \ldots \mathbf{p}_N \mathbf{q}_1 \ldots \mathbf{q}_N)$ denotes the phase of the system, the variable $\alpha$ is defined by requiring the internal energy $H_0(\mathbf{x}) = \sum_{j=1}^N \mathbf{p}_j^2/2m + \Phi(\mathbf{q}_1...\mathbf{q}_N)$ to be a constant of motion, where $\Phi(\mathbf{q}_1 \ldots \mathbf{q}_N) = \sum_{i<j} \varphi(|\mathbf{q}_i - \mathbf{q}_j|)$, and $\alpha$ can be easily computed [15]. The model is studied in [14], [15], [18] with periodic (Lees-Edwards) boundary conditions, which we replace here by simpler ones: horizontally periodic boundary conditions and vertically elastic wall reflections, [3].

For a dynamical description of this system one can suppose that the total (peculiar) momentum $P_\parallel$ and the center of mass position $X_\parallel$ in the shear direction vanish. If the observations are made in discrete time by observing subsequent particle collisions, then, with the constant energy, the system can be described in a phase space $\mathcal{C}$ of $4N - 4$ dimensions. The Liouville distribution can be projected on $\mathcal{C}$ giving a distribution $\bar{\mu}_0$ (to which we refer with the same name), but $\bar{\mu}_0$ is stationary only for $\gamma = 0$. The evolution will be a map $S$ related to the solution operator $t \to S_t\mathbf{x}$ of (1) and to the time $t(\mathbf{x})$ elapsing between a timing event $\mathbf{x} \in \mathcal{C}$ and the next: $S\mathbf{x} = S_{t(\mathbf{x})}\mathbf{x}$. The phase space volume variation rate, which is also the entropy production rate for this system [23,11] is the divergence of the r.h.s. of Eq.(1) and equals $(2N - 2)\sigma$, with $\sigma$ close to $\alpha$; in fact one finds, after a brief computation: $\sigma = \alpha + \gamma \frac{\sum_j p_{xj}p_{yj}}{(N-1)\sum_j \mathbf{p}_j^2} = \alpha + \gamma O(N^{-1})$. The average entropy production rate is $(2N - 2)\langle\sigma\rangle$ where the brackets denote a (forward) time average over an infinite time.

The many particle statistical mechanical system in Eq.(1) exhibits the following features:
(A) Dissipation: $\langle\sigma\rangle > 0$, [4];
(B) Time reversal invariance: the map $i: (x, y, p_x, p_y) \to (x, -y, -p_x, p_y)$ is such that if

$t \to \mathbf{x}(t)$ is a solution of (1), so is $i(\mathbf{x}(-t))$; [16,15].

We note that (A) implies the existence of an invariant set $A$, which we will call the *attractor*, of zero Liouville probability (if $\gamma \neq 0$) but probability 1 with respect to the asymptotic statistics $\bar{\mu}$ of the motions with random initial data (with the Liouville distribution), so that $\bar{\mu}(A) = 1$. Property (B) implies a direct relation between the statistical properties of the forward ($t \to \infty$) and backward ($t \to -\infty$) motions, although described in general by different statistics: e.g. they have the same set of Lyapunov exponents.

To apply Ruelle's principle, in the spirit discussed above, we assume that, in view of the particle collisions, the behavior on $A$ is as if $A$ were hyperbolic. There may well be corrections to this: proceeding by ignoring such possibilities is our interpretation of (R). In practice this means that we suppose that the *corrections become negligible as $N \to \infty$*. [5] The strongest form of the hyperbolicity assumption is:

(C) (a) at every point $\mathbf{x}$ of $\mathcal{C}$ one can define stable and unstable manifolds $W^s_{\mathbf{x}}, W^u_{\mathbf{x}}$ *dense* in $\mathcal{C}$, *transversal* and *covariant* (i.e. they form an angle bounded away from $0, \pi$ when they cross and $SW^\beta_{\mathbf{x}} = W^\beta_{S\mathbf{x}}$, $\beta = u, s$). (b) the tangent planes to $W^\beta_{\mathbf{x}}$ vary continuously with $\mathbf{x}$. (c) line elements at $\mathbf{x}$ are uniformly reduced in length by a factor at least $Ce^{-\lambda n}$ under the action of $S^n$ if tangent to $W^s_{\mathbf{x}}$, or under the action of $S^{-n}$ if tangent to $W^u_{\mathbf{x}}$, for $n \geq 0$ with $C, \lambda > 0$; (d) if the sign of $n$ is changed corresponding properties hold with expansions replacing contractions.

Such a system has been considered in [29] (Anosov system) and it can be shown that it admits a distribution $\bar{\mu}$ describing the statistics of random data chosen initially with the Liouville distribution and that $\bar{\mu}$ verifies (R), being uniquely determined by the property stated in (R). [6]

We note that (B),(C) imply that there are $2N - 2$ positive Lyapunov exponents for $S$ ($S^{-1}$) and that $W^u_{\mathbf{x}}$ ($W^s_{\mathbf{x}}$) have the symmetry property $W^s_{\mathbf{x}} = iW^u_{i\mathbf{x}}$. Furthermore $A$ (and $iA$) are dense in $\mathcal{C}$, although we anticipate, from the pairing property in [14] and the expected smoothness of the Lyapunov spectrum [22], that the Kaplan Yorke fractal dimension of $A$, [17], is $\gamma^2 O(N)$ smaller than the dimension of $\mathcal{C}$ (i.e. "dimensional reduction" occurs).

Most results deriving from (C) really come from the fact that (C) implies the existence of *Markov partitions* (MP), which permit us to give a rather simple description of the distribution $\bar{\mu}$ (the SRB, see Eq.(3) below), upon which all our deductions are based. The MP are *tilings* $\mathcal{E}$ of $\mathcal{C}$ with suitable small sets $E$ called "parallelograms", naturally related to the stable and unstable manifolds (with their "axes" parallel to the stable and to the unstable manifolds), see [10], [26] for a complete description. A property that always holds is the time invariance of the MP: if $\mathcal{E}$ is a MP also the partition $S\mathcal{E}$ obtained by evolving in time the elements of $\mathcal{E}$ is such. A key property of MP is that the partition obtained by considering the intersections of all the tiles of two MP is still a MP. Hence, given a MP $\mathcal{E}$, one can construct other much finer ones: a typical method is to consider the partitions $S^q\mathcal{E}$ and to define the partition $\mathcal{E}_T$ obtained by intersecting all the tiles of all the partitions $S^q\mathcal{E}$ with $-T < q < T$. The parallelograms of the finer partitions may become as small as desired by taking $T$ large (by the hyperbolicity). The description of the statistics $\bar{\mu}$ of the Liouville distribution, in terms of the family of finer partitions $\mathcal{E}_T$, associated with $\mathcal{E}$, is done as follows:

1) Given two times $\tau$ and $T$, let $E_j$ be the parallelograms of $\mathcal{E}_T$ labeled by $j$, and let $\mathbf{x}_j$ be a point in $E_j$ (arbitrarily fixed); define a probability distribution $\bar{\mu}_{\tau,T}$ by attributing to each $E_j \in \mathcal{E}_T$ a weight:

$$\bar{\Lambda}_{u,\tau}^{-1}(\mathbf{x}_j) = \prod_{k=-\tau/2}^{\tau/2-1} \Lambda_u^{-1}(S^k \mathbf{x}_j) \tag{2}$$

where $\Lambda_u(\mathbf{x})$ is the absolute value of the determinant of the jacobian matrix of $S$ as a map of $W_{\mathbf{x}}^u$ to $W_{S\mathbf{x}}^u$, so that the weight $\bar{\Lambda}_u^{-1}(\mathbf{x}_j)$ is the inverse of the expansion coefficient of the map $S^\tau$ evaluated at $S^{-\tau/2}\mathbf{x}_j$, i.e. at the initial point of a motion spending half of the time "before" reaching $\mathbf{x}_j$ and half "after". For later use we define corresponding quantities associated with the stable manifold, denoted by $\Lambda_s(\mathbf{x})$ and $\bar{\Lambda}_{s,\tau}(\mathbf{x})$.

2) After normalization the above weigths define a distribution $\bar{\mu}_{\tau,T}$ by requiring that the average of any (smooth) function $G$ with respect to the distribution $\bar{\mu}_{\tau,T}$ will be:

$$\int_{\mathcal{C}} \bar{\mu}_{\tau,T}(d\mathbf{x})G(\mathbf{x}) = \frac{\sum_j \bar{\Lambda}_{u,\tau}^{-1}(\mathbf{x}_j)G(\mathbf{x}_j)}{\sum_j \bar{\Lambda}_{u,\tau}^{-1}(\mathbf{x}_j)} \qquad (3)$$

3) Consider the limit as $\tau, T \to \infty$, with $T \to \infty$ fast enough compared to $\tau$ so that in each parallelogram the weights in Eq.(3) have a small relative variation.

It follows then from Sinai's work [29] and [10], [26] that *the limit exists and is the statistics $\bar{\mu}$ of the Liouville distribution* [21] and that $\bar{\mu}$ does indeed verify (R). The $E_j$ may be interpreted as the *dynamical states* of [15]. The $\bar{\mu}_{\tau,T}$ in Eq.(3) plays the role of the finite volume Gibbs distributions in statistical mechanics.

We now turn to the numerical experiment in [15]:

(i) the main results concern properties of the entropy production over a generic time interval $\tau t_0$, during which the trajectory moves between $S^{-\tau/2}\mathbf{x}$ and $S^{\tau/2}\mathbf{x}$, if $t_0$ is the average time interval between two successive collisions, [7] ($\mathbf{x}$ is the middle point of this trajectory segment). The entropy production is defined here, following [23], by $t_0 \sum_{-\tau/2}^{\tau/2-1}(2N-2)\sigma(S^j \mathbf{x})$. It will be convenient to rewrite the latter expression by setting $(2N-2)\tau t_0 \langle \sigma \rangle_\tau(\mathbf{x}) = (2N-2)\tau t_0 \langle \sigma \rangle a_\tau(\mathbf{x})$, where $\langle \sigma \rangle_\tau$ denotes the time average of the entropy production rate $\sigma$ over the time interval $\tau t_0$. This defines a fluctuation variable $a_\tau(\mathbf{x})$ with (forward) average equal to 1 (because the infinite time average of $\langle \sigma \rangle_\tau$ is the above defined $\langle \sigma \rangle$, see (A)). The exponential $\mathcal{F}_\tau(\mathbf{x})$ of the entropy production is the reciprocal of the phase space contraction itself.

(ii) For such an observable, *which is strongly $\tau$ dependent*, it may be doubtful to use $\bar{\mu}_{\tau,T}$ to estimate the probabilities relative to $\bar{\mu}$, even if $\tau, T$ are very large. [8] However, it is possible to prove, if (C) holds, ([21]), that not only the average of $\mathcal{F}_\tau$ but *even the probabilities of the various values of $\mathcal{F}_\tau$ can be computed by using $\bar{\mu}_{\tau,T}$* of Eq.(3). The error on the probability attributed to each individual $E_j$ would then consist of a factor bounded, above and below, by $\tau$-*independent* positive constant.

In the computer experiment [15] one measures the entropy production $\log \mathcal{F}_\tau(\mathbf{x})$, as seen on a stretch of time $\tau$ short compared to the experiment duration $T$, repeatedly $T/\tau$ times (see also [18]). This amounts to studying the $\bar{\mu}$ distribution of the fluctuation $a_\tau(\mathbf{x})$, in (i).

The main experimental result corresponds in the present context to the computation of the logarithm of the ratio of the probability that $a_\tau(\mathbf{x}) = p$ to that of $a_\tau(\mathbf{x}) = -p$. The result, fig. 2 of [15], is that this quantity as a function of $p$ is a precise straight line with slope $2N\tau t_0 \langle \sigma \rangle$ for $\tau$ large.

We now use the very formal expression of $\bar{\mu}$ defined via Eq.(3) to study some statistical properties of $\mathcal{F}_\tau(\mathbf{x})$ and compare the result with [15]. We argue that the ratio of the probability of $a_\tau(\mathbf{x}) \in [p, p+dp]$ to that of $a_\tau(\mathbf{x}) \in [-p, -p+dp]$ is, using the notations and the approximation to $\bar{\mu}$ in Eq.(3):

$$\frac{\sum_{j,\, a_\tau(\mathbf{x}_j)=p} \bar{\Lambda}^{-1}_{u,\tau}(\mathbf{x}_j)}{\sum_{j,\, a_\tau(\mathbf{x}_j)=-p} \bar{\Lambda}^{-1}_{u,\tau}(\mathbf{x}_j)} \tag{4}$$

We evaluate Eq.(4) by establishing a one to one correspondence between addends in the numerator and in the denominator, aiming at showing that corresponding (i.e. with the same $j$) addends have a *constant ratio* which will, therefore, be the value of the ratio in Eq.(4). This is made possible by the time reversal symmetry which is the extra information we have with respect to [29], [10], [26]. In fact the time reversal symmetry (B) can be shown to imply that with $\mathcal{E}$ also $i\mathcal{E}$ is a MP. Since the intersections of MP are still MP we can assume that $\mathcal{E}$, hence also $\mathcal{E}_T$, will be time reversal symmetric, i.e. that for each $j$ there is a $j'$ such that $iE_j = E_{j'}$. By using the identities $S^{-\tau}(S^\tau \mathbf{x}) = \mathbf{x}$, and $S^{-\tau}(iS^{-\tau}\mathbf{x}) = i\mathbf{x}$ (time reversal) and $iW^u_\mathbf{x} = W^s_{i\mathbf{x}}$, one can deduce, with $\bar{\Lambda}_{s,\tau}(\mathbf{x})$ defined after Eq.(2): $a_\tau(\mathbf{x}) = -a_\tau(i\mathbf{x})$ and $\bar{\Lambda}_{u,\tau}(i\mathbf{x}) = \bar{\Lambda}_{s,\tau}(\mathbf{x})^{-1}$. The ratio Eq.(4) can therefore be rewritten as:

$$\frac{\sum_{E_j, a_\tau(\mathbf{x}_j)=p} \bar{\Lambda}^{-1}_{u,\tau}(\mathbf{x}_j)}{\sum_{E_j, a_\tau(\mathbf{x}_j)=-p} \bar{\Lambda}^{-1}_{u,\tau}(\mathbf{x}_j)} \equiv \frac{\sum_{E_j, a_\tau(\mathbf{x}_j)=p} \bar{\Lambda}^{-1}_{u,\tau}(\mathbf{x}_j)}{\sum_{E_j, a_\tau(\mathbf{x}_j)=p} \bar{\Lambda}_{s,\tau}(\mathbf{x}_j)} \tag{5}$$

Then the ratios between corresponding terms in Eq.(5) are equal to $F_\tau(\mathbf{x}_j) \equiv \bar{\Lambda}^{-1}_{u,\tau}(\mathbf{x}_j) \cdot \bar{\Lambda}^{-1}_{s,\tau}(\mathbf{x}_j)$. This is almost $\mathcal{F}_\tau(\mathbf{x}_j) \equiv e^{a_\tau(\mathbf{x}_j)(2N-2)\langle\sigma\rangle t_0 \tau}$ *which is $j$–independent* (because $a_\tau(\mathbf{x}_j) \equiv p$). In fact, the latter is the reciprocal of the determinant of the jacobian matrix of $S$, i.e. the reciprocal of the total phase space volume variation, while $F_\tau(\mathbf{x}_j)$ is only the reciprocal of the product of the variations of two surface elements tangent to the stable and to the unstable manifold in $\mathbf{x}_j$. Hence $F_\tau(\mathbf{x})$ and $\mathcal{F}_\tau(\mathbf{x})$ differ by a factor related to the angles between the manifolds at $S^{-\tau/2}\mathbf{x}$ and at $S^{\tau/2}\mathbf{x}$. But assumption (C) implies transversality of their intersections, so that the ratio between $\mathcal{F}_\tau(\mathbf{x})$ and $F_\tau(\mathbf{x})$ is bounded away from 0 and $+\infty$ by $(\mathbf{x},\tau)$–independent constants. *Therefore the ratio Eq.(4) is equal to $e^{p(2N-2)\langle\sigma\rangle t_0\tau}$ as well as to that found in* [15], up to a factor bounded above and below by a $(\tau,p)$ independent constant, i.e. to leading order as $\tau\to\infty$.

We note that this result can be considered as a large deviation result (both in $N$ and $t_0\tau$): its peculiarity is the linearity in $p$. On general grounds one might expect that the deviations of $\frac{1}{(2N-2)t_0\tau}\log\mathcal{F}_\tau(\mathbf{x})$, or of $p$, from its average value 1 have a probability density $\pi(p)$ expressible in terms of some "free energy" function $\zeta(p)$ as $\pi(p)=e^{-\zeta(p)\tau t_0}$. Here $\zeta(p)$ describes the multifractal nature of the observable $\mathcal{F}_\tau$: a "non multifractal" distribution would correspond to a delta function distribution $e^{\zeta(p)\tau t_0}\equiv\delta(p-1)$. Noting that $\varepsilon(p)\equiv\zeta(p)-\zeta(-p)=\frac{1}{\tau t_o}\log\pi(p)/\pi(-p)$ is an odd function of $p$, we expect that $\varepsilon(p)=c\langle\sigma\rangle(p+s_3p^3+s_5p^5+\ldots)$ with $c>0$ and $s_j\neq 0$, since there is no reason, a priori, to expect a "simple" (i.e. with linear odd part) multifractal [9]. *Thus p–linearity (i.e. $s_j\equiv 0$) is a key test of the theory* and a quite unexpected result from the latter viewpoint. Recall, however, that as noted in remark (ii) after Eq.(3) the exponent $(2N-2)t_0\tau\langle\sigma\rangle p$ is correct up to terms of $O(1)$ in $\tau$ (i.e. deviations at small $p$ must be expected). This shows that Ruelle's principle can indeed be tested on many particle systems in statistical mechanics and in fact is in agreement with the computer results obtained in [15]. The same theory would also apply to other models, like the ones in [11] or in the time reversible case in [12] for which, to date, no corresponding experimental results exist, yet.

A similar test might even apply to fluid dynamics, for which (R) was originally devised. In fact the above linear multifractality (of the odd part of $\zeta(p)$) might be observable in the high turbulence regime in fluid mechanics at least in numerical experiments for models in which no friction acts in the inertial range. The fluid is then described in the inertial range by the reversible Euler equations, while the dissipation below the Kolmogorov scale can be modeled by a gaussian thermostat thus making the equations describing the system analogous to Eq.(1) (see also [23]).

*Aknowledgements:* We are grateful to J. L. Lebowitz, G. L. Eyink, Y.Sinai for explaining their work [11] [12] and for important comments. E.G.D.C. acknowledges support from DOE-grant DE-FG02-88-ER13847 and G.G acknowledges partial financial support from Rutgers U., Rockefeller U. and CNR-GNFM.

# REFERENCES

[1] (R) cannot be proved in general. Motion hyperbolicity is usually unknown.

[2] A task possible in the $N = 1$ case considered in [11] but essentially beyond our capabilities in slightly more general systems, certainly in the non linear regime.

[3] Modeled by a strong wall repulsion.

[4] Positivity rests on numerical evidence [14].

[5] This is usually also assumed in dealing with the ergodic hypothesis, or more generally with principles which cannot be assumed to be independent from the basic laws of motion.

[6] The assumptions (C) may just be too strong and/or difficult to verify for the model in Eq.(1). Their strength can be seen from the fact that they imply (R) above, [29], [26], and (hence) the ergodic hypothesis if $\gamma = 0$, [29], [26]. The density in $\mathcal{C}$ could be replaced by the requirement of density in $A$ alone; this would be more general without affecting our conclusions. Furthermore our main conclusions could still be reached under far weaker assumptions, as we think that the consequences of (C) relevant for our analysis naturally follow in the frame of the theory of singular hyperbolic systems of [24].

[7] We identify this with the actual time interval between them, neglecting its fluctuations (which could be easily taken into account leading to the same end result).

[8] This is just as in the Ising model where one cannot compute correctly the thermodynamic limit average of the magnetization (or the average of any extensive quantity) for a finite volume $\tau$ by using the finite volume Gibbs distribution *with the same volume $\tau$* (without incurring large, size dependent, errors). One *must* use a larger volume $\theta \gg \tau$, except in one dimension where it is well known [26] that the probability distribution of the magnetization would be off "only" by a factor bounded above and below for each spin configuration by a $\tau$–independent constant. Our case is very similar, as follows

from the thermodynamic formalism [29], [26], (see [21] for a treatment in the frame of the present problem).